\documentclass[onecolumn,nofootinbib,preprintnumbers]{revtex4-2}
\usepackage{amssymb}
\usepackage{amssymb}
\usepackage{mathrsfs}
\usepackage{float}
\usepackage{bm}
\usepackage{amsmath}
\usepackage{mathdots}
\usepackage{graphicx}
\usepackage{subfigure}
\usepackage{array}
\usepackage{lipsum}
\usepackage{color}
\usepackage{hyperref}
\usepackage{ulem}
\begin{document}
	
\title{Comment on ``Quantum trapping and rotational self-alignment in triangular Casimir microcavities''}

\newcommand*{\SDUST}{Department of Physics and Institute for Theoretical Physics,\\Shandong University of Science and Technology, Qingdao, Shandong 266590, China}\affiliation{\SDUST}

\author{Zhentao Zhang}\email{zhangzt@sdust.edu.cn}\affiliation{\SDUST}

\begin{abstract}
	Küçüköz \textit{et al}., Sci. Adv. \textbf{10}, eadn1825 (2024) reported an experiment, which shows that conducting plates in the misaligned system separated by a liquid may experience transverse interactions at small separations. We point out that the effects might have been examined in detail in theoretical studies, New J. Phys. \textbf{24}, 113036 (2022) and Phys. Lett. A \textbf{495}, 129304 (2024).
\end{abstract}

\maketitle

Recently, Küçüköz \textit{et al.} reported an experiment~\cite{SA}, which shows that gold plates in the misaligned system separated by a liquid can experience a torque and a lateral force at small separations. We would like to point out that the effects might have been theoretically examined in detail in Refs.~\cite{Zhang1,Zhang2}. The studies focused on the misaligned system of uncharged parallel layers, and discussed a tangential Casimir force\footnote{A force that is parallel to flat surfaces is called tangential force in these studies.} and a Casimir torque experienced by \textit{flat} plates made of isotropic media, see also~\cite{Wagner}. To show the presence of these interactions by the analytic formulation, Ref.~\cite{Zhang1} conducted a detailed, necessary investigation on the influence of the edge effect. 

It may be easy to observe that~\cite{Zhang1,Zhang2} the properties of the tangential force and the torque discussed by the studies are different to the lateral Casimir force and the Casimir torques induced by periodically corrugated surfaces or anisotropic materials. To facilitate experimental investigation of these effects, Refs.~\cite{Zhang1,Zhang2} studied various practical examples. For instance, Ref.~\cite{Zhang2} provided the analytic prediction for a torque experienced by the gold plates in the misaligned system, which would be reliable for nearly all of the parameter space except for the small-angle limit ($\theta\rightarrow0$), see, e.g., Figs. 7 and 8 therein.

Hence, it might be appropriate to say that the studies may be relevant theoretical background to the experiment in Ref.~\cite{SA}. Incidentally, we may remark that Ref.~\cite{Na} gave an estimation of a stability probably ensured by a constant lateral Casimir-Electrostatic force between two charged plates in the marginally-misaligned~\cite{Zhang1} system, without studying the potentially overwhelming influences of the edge effects, which, however, could also have attracted theoretical investigations on the existence of the tangential Casimir force back then. 

We end the comment by noting that, to verify the predicted effects conclusively, measurement of the tangential force and the torque by precisely controllable experimental schemes will be needed.

\vspace{0.1cm}

\textit{Acknowledgment}. We thank T. O. Shegai for the communication.

\end{document}